\documentclass[%
 reprint,
 superscriptaddress,
 amsmath,amssymb,
 prx,
 aps,
 floatfix,
 longbibliography
]{revtex4-2}

\usepackage{graphicx}
\usepackage{dcolumn}
\usepackage{bm}
\usepackage{siunitx}
\usepackage{hyperref}
\usepackage{xcolor}
\usepackage[utf8]{inputenc}
\usepackage[T1]{fontenc}
\usepackage{soul}
\usepackage{upgreek}

\usepackage[left]{lineno}
\modulolinenumbers[5]
\renewcommand{\makeLineNumber}{\llap{\linenumberfont\rlap{\LineNumber}\hspace{0.45cm}}}
\renewcommand{\linenumberfont}{\tiny}

\newcommand{\Ej}{$E_{\rm J}$}
\newcommand{\Vj}{$V_{\rm j}$}

\begin{document}

\preprint{APS/123-QED}

\title{A gate-tunable, field-compatible fluxonium}

\author{Marta Pita-Vidal}
\affiliation{QuTech, Delft University of Technology, 2628 CJ Delft, The Netherlands.}
\affiliation{Kavli Institute for Nanoscience, Delft University of Technology, 2628 CJ Delft, The Netherlands.}

\author{Arno Bargerbos}
\affiliation{QuTech, Delft University of Technology, 2628 CJ Delft, The Netherlands.}
\affiliation{Kavli Institute for Nanoscience, Delft University of Technology, 2628 CJ Delft, The Netherlands.}

\author{Chung-Kai Yang}
\affiliation{Microsoft Quantum Lab Delft, 2628 CJ Delft, The Netherlands.}

\author{David J. van Woerkom}
\affiliation{Microsoft Quantum Lab Delft, 2628 CJ Delft, The Netherlands.}

\author{Wolfgang Pfaff}
\affiliation{Microsoft Quantum Lab Delft, 2628 CJ Delft, The Netherlands.}

\author{Nadia Haider}
\affiliation{QuTech, Delft University of Technology, 2628 CJ Delft, The Netherlands.}
\affiliation{Netherlands Organisation for Applied Scientific Research (TNO), 2628 CK Delft, The Netherlands}

\author{Peter Krogstrup}
\affiliation{Center for Quantum Devices, Niels Bohr Institute, University of Copenhagen, 2100 Copenhagen, Denmark and  Microsoft Quantum Materials Lab Copenhagen, 2800 Lyngby, Denmark.}

\author{Leo P. Kouwenhoven}
\affiliation{QuTech, Delft University of Technology, 2628 CJ Delft, The Netherlands.}
\affiliation{Microsoft Quantum Lab Delft, 2628 CJ Delft, The Netherlands.}

\author{Gijs de Lange}
\affiliation{Microsoft Quantum Lab Delft, 2628 CJ Delft, The Netherlands.}

\author{Angela Kou}
\affiliation{Microsoft Quantum Lab Delft, 2628 CJ Delft, The Netherlands.}


\date{\today}

\begin{abstract}
Hybrid superconducting circuits, which integrate non-superconducting elements into a circuit quantum electrodynamics (cQED) architecture, expand the possible applications of cQED. Building hybrid circuits that work in large magnetic fields presents even further possibilities such as the probing of spin-polarized Andreev bound states and the investigation of topological superconductivity. Here we present a magnetic-field compatible hybrid fluxonium with an electrostatically-tuned semiconducting nanowire as its non-linear element. We operate the fluxonium in magnetic fields up to 1T and use it to observe the $\varphi_0$-Josephson effect. This combination of gate-tunability and field-compatibility opens avenues for the control of spin-polarized phenomena using superconducting circuits and enables the use of the fluxonium as a readout device for topological qubits.

\end{abstract}

\maketitle



%
\section{Introduction}
Circuit quantum electrodynamics, where photons are coherently coupled to artificial atoms built with superconducting circuits, has enabled the investigation and control of macroscopic quantum-mechanical phenomena in superconductors \cite{Blais2004, Chiorescu2004, Wallraff2004}. Recently, hybrid circuits incorporating semiconducting nanowires \cite{deLange2015, Larsen2015,  Luthi2018, Hays2018} and other electrostatically-gateable elements \cite{Mi2017, Casparis2018, Kroll2018, Wang2019} into superconducting circuits have broadened the scope of cQED to probing mesoscopic superconductivity \cite{deLange2015, Larsen2015, Hays2018, Hays2019}. Further extending the capabilities of hybrid circuits to work in magnetic fields presents the intriguing possibility of insights into \textit{topological} superconductivity \cite{Zazunov2014, Vayrynen2015, vanHeck2017, Yavilberg2019,Hassler2011, Hyart2013,  Ginossar2014, Stenger2019}. 

Topological superconductivity, which has garnered much theoretical and experimental interest lately  \cite{Kitaev2001, [{}][{ and references therein.}]{Lutchyn2018}, [{}][{ and references therein.}]{Sato2017}, [{}][{ and references therein.}]{Sarma2015}}, results from the interplay between magnetism and superconductivity. In superconductor-proximitized semiconductors exposed to a large magnetic field, emergent quasiparticle states known as Majorana zero modes (MZM) can form. Majorana zero modes are predicted to be robust to local perturbations and could thus serve as long-lived qubits. Several groups have proposed using superconducting circuits to both probe and control MZMs \cite{Vayrynen2015, Hassler2011, Hyart2013, Pekker2013, Stenger2019}. All of the above-mentioned proposals, however, require the operation of a flux-based superconducting circuit in large magnetic fields. 

Additionally, the behavior of superconductor-proximitized semiconductors exposed to a large magnetic field in the trivial phase remains an active field of research. Most experiments investigating the behavior of Andreev bound states (ABS) in a magnetic field have been focused on the magnetic-field dependence of the switching current \cite{Paajaste2015, Zuo2017, Hart2019} with the exceptions of references~\cite{vanWoerkom2017} and~\cite{Tosi2019}, which performed ABS spectroscopy up to fields of \SI{300}{mT} and \SI{11}{mT} respectively. Energies below \SI{20}{\upmu eV}, which are interesting for high-transparency ABSs and MZMs, were, however, not accessible. A fluxonium that works in high magnetic fields and with semiconducting weak-links would thus be an extremely useful tool both for investigating the magnetic field behavior of ABSs and also for coupling to topological superconductors. First, the fluxonium ground-to-excited state transitions are typically accessible at microwave frequencies over the entire flux range from 0 to $\pi$. One therefore gains access to the full energy-phase relations of the junction. Second, the Josephson energy of the junction can be known extremely precisely (to $<$ \SI{0.4}{\upmu eV}) and over multiple decades, which allows for the mapping of the characteristic energy of the semiconducting weak-link. Finally, the fluxonium is also sensitive to quasiparticle poisoning events in individual ABS in the junction since poisoning of individual ABSs would lead to additional spectral lines. Knowledge of these rates would greatly aid the design of Andreev-based qubits \cite{Janvier2015, Hays2019}.

Moreover, the unique parameter regime of the fluxonium makes it particularly suited to detecting and controlling MZMs. A switch in the parity of MZMs coupled to the fluxonium circuit corresponds to a switch in the direction of the persistent current flowing in the fluxonium circuit. Parity switches of the MZMs would thus result in the observation of two copies of the fluxonium spectrum \cite{Pekker2013}. In addition, the presence of MZMs coupled to the fluxonium changes the periodicity of its spectrum from $2\pi$ to $4\pi$. In these proposals, the linewidth of the fluxonium transition determines the sensitivity with which the fluxonium could determine the presence of MZMs. Beyond its detection capabilities, multiple fluxonium devices coupled to MZMs could also implement braiding operations on the MZMs \cite{Stenger2019}.

Here we have realized a hybrid fluxonium incorporating an Al-proximited nanowire that operates up to 1T. We build upon recent work proximitizing semiconducting nanowires to incorporate a magnetic-field compatible, electrostatically tunable weak-link junction into the fluxonium \cite{Krogstrup2015}. The presented fluxonium device also has a gradiometric design and is composed of NbTiN for further field-compatibility. We demonstrate in-situ gate tunability of the fluxonium Josephson energy over more than a decade. We then operate the fluxonium in fields up to 1T and map out the dependence of the fluxonium Josephson energy as a function of field. In all regimes of magnetic field and gate voltage, we observe excellent agreement between the data and a theoretical model based on a simple Hamiltonian with a few degrees of freedom. Finally, we also demonstrate the utility of the fluxonium as a probe of mesoscopic superconductivity in magnetic fields by using the measured energy-phase relation of the junction to observe the $\varphi_0$-Josephson effect. The ability to observe the fluxonium spectrum over wide ranges in gate voltage and magnetic fields establishes the hybrid fluxonium as a novel superconducting circuit for exploring superconducting phenomena in a magnetic field and as realistic readout platform for MZMs.  

\begin{figure}
    \centering
    \includegraphics[scale=1.0]{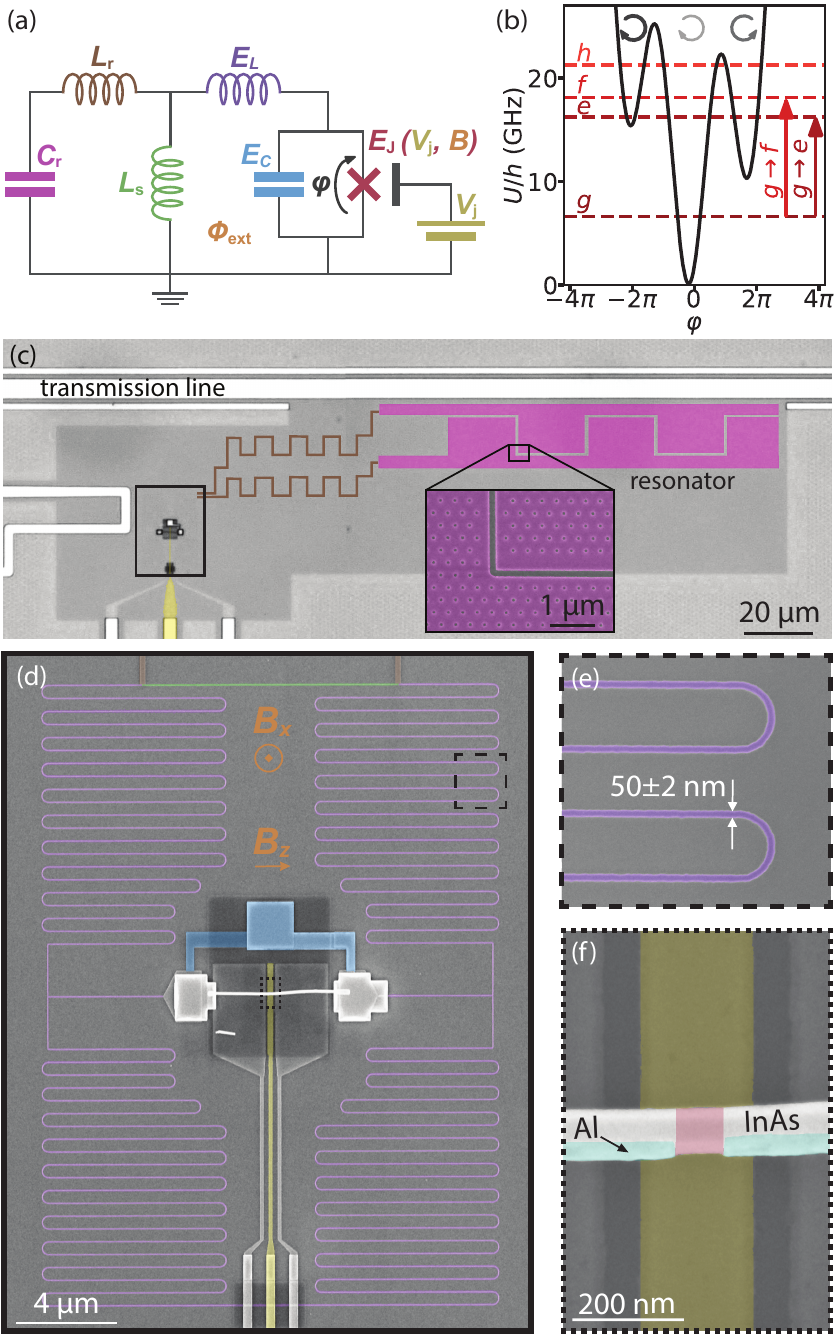}
    \caption{{Nanowire fluxonium.} (a) Circuit model. The fluxonium is composed of a Josephson junction shunted by an inductor and a capacitor, which are characterized by the energies \Ej, $E_L$, and $E_C$, respectively. The value of \Ej~ depends on the external magnetic field, $B_z$, and the gate voltage, $V_{\rm j}$. An readout resonator (constituted by $L_{\rm r}$ and $C_{\rm r}$) is coupled to the fluxonium via a shared inductance $L_{\rm s}$. (b) Potential of the fluxonium (black) versus the phase difference across the junction, $\varphi$, at $\varphi_{\rm ext}=0.2\pi$. The lowest eigenenergies are indicated with dashed horizontal lines. Red arrows indicate transitions starting from the ground state. Circular gray arrows represent the amplitude and direction of the persistent-current states associated with different potential wells. (c) False-colored optical image showing the transmission line and the resonator, with capacitive and inductive elements shaded in pink and brown respectively. Inset, scanning electron microscope (SEM) image of the resonator's capacitive plates. (d) SEM image of a lithographically similar fluxonium to Device A, corresponding to the area indicated by the box in (c). The NbTiN superinductor (purple, enlarged in (e)), the shared inductance section (green), the parallel plate capacitor (blue) and the nanowire junction (red, enlarged in (f)), correspond, respectively, to the $E_L$, $L_{\rm s}$, $E_C$ and \Ej~elements in (a). The out-of-plane component of $B$, $B_x$, tunes the external magnetic flux $\Phi_{\rm ext}$. $B_z$ is the component parallel to the wire.}
    \label{fig:1}
\end{figure}

\section{Magnetic-field compatible materials and design}
Building a fluxonium compatible with the application of a magnetic field presents multiple challenges. The first challenge is reaching the fluxonium regime using magnetic-field compatible materials. The fluxonium consists of a Josephson junction with Josephson energy \Ej~in parallel with a linear superinductor \cite{Manucharyan2012}, with inductive energy $E_L$, and a capacitor characterized by the energy $E_C$, as shown in Fig.~\ref{fig:1}(a). The fluxonium regime ($E_L<E_C<E_{\rm J}$) is achieved by shunting the junction with a large inductance. This parameter regime results in the eigenstates of the fluxonium being composed of superpositions of persistent currents in multiple directions (Fig.~\ref{fig:1}(b)). Additionally, the chosen fluxonium parameters result in a spectrum that is observable over the entire range of applied external flux. Superconductor-insulator-superconductor (SIS) Josephson junction arrays, commonly used to implement the fluxonium superinductance \cite{Manucharyan2009, Manucharyan2012, Kalashnikov2020} can not be used since they are incompatible with large magnetic fields. Recent work on magnetic-field compatible materials with a large kinetic inductance such as granular aluminium \cite{Rotzinger2016, Grunhaupt2018, Maleeva2018, Grunhaupt2019} and NbTiN \cite{Annunziata2010, Samkharadze2016, Hazard2019} has presented a path to meeting the stringent requirements of the fluxonium superinductance. The magnetic-field compatible fluxonium device is shown in Figs~\ref{fig:1}(c-f). All circuit elements except for the junction are fabricated using NbTiN, which has been demonstrated to have critical fields exceeding \SI{9}{T} and inductances exceeding \SI{75}{pH\per\Box} \cite{Samkharadze2016}. Here we define the fluxonium by etching a \SI{9}{nm}-thick sputtered NbTiN film, which has a kinetic inductance of \SI{41}{pH\per\Box}. The superinductance of the fluxonium is made with a \SI{50}{nm}-wide NbTiN meander (Fig.~\ref{fig:1}(e)) in order to maximize the inductance of the device while minimizing spurious capacitances to ground. This design realizes an inductance of $\sim \SI{100}{nH}$. The small width of the meanders additionally suppresses the emergence of lossy vortices due to out-of-plane fields, $B_{x}$, up to tens of mT \cite{Stan2004}. We further mitigate the effects of these vortices by introducing vortex-pinning holes (inset, Fig.~\ref{fig:1}(c)) into the capacitor of the fluxonium readout resonator and the ground plane.  

In addition to being composed of magnetic-field compatible materials, for use as a detector, the fluxonium must also maintain its narrow linewidth during the application of a magnetic field. The application of a magnetic field precludes the possibility of using the magnetic shielding necessary for limiting flux noise in flux-based superconducting circuits. We address this challenge by building a gradiometric superinductance as shown in Fig.~\ref{fig:1}(d). Equal fluxes through each of the two loops generate equal currents that are canceled at the junction, rendering the fluxonium insensitive to flux noise due to sources larger than the fluxonium device.

Finally, a SIS Josephson junction  made of Al and AlOx, which has been used in all previously reported fluxonium devices, can not be used as the \Ej~element here due to its incompatibility with magnetic field. We build a magnetic-field compatible junction by incorporating a semiconducting InAs nanowire proximitized by an epitaxially-grown \SI{6}{nm}-thick aluminum layer (Fig.~\ref{fig:1}(f)) \cite{Krogstrup2015} into the fluxonium. The small thickness of the aluminum shell makes it resilient to magnetic fields along the wire of more than \SI{1}{T} \cite{Krogstrup2015}. Similar resilience to magnetic fields could also be attained using weak links of disordered superconducting materials, such as indium-oxide \cite{Astafiev2012} or granular aluminum \cite{Winkel2019}, as the Josephson junction. The nanowire is deterministically deposited on top of the pre-patterned leads of the inductor and capacitor using a micromanipulator. The Josephson junction is then defined by etching away an Al section of $\sim\SI{80}{nm}$ on top of the junction gate. This small junction, however, does not provide a large enough capacitance to achieve the fluxonium regime for typical \Ej~values in nanowire junctions. We thus add a parallel plate capacitor (blue in Fig.~\ref{fig:1}(d)) to decrease $E_C$ and achieve its required value for the fluxonium. The fluxonium capacitor consists of two square NbTiN plates sandwiching a \SI{29}{nm}-thick SiN dielectric. We note that this fluxonium design is flexible enough to incorporate any semiconducting material as its small junction. 

\section{Fluxonium Spectroscopy}

\begin{figure}[t!]
    \centering
    \includegraphics[scale=1.0]{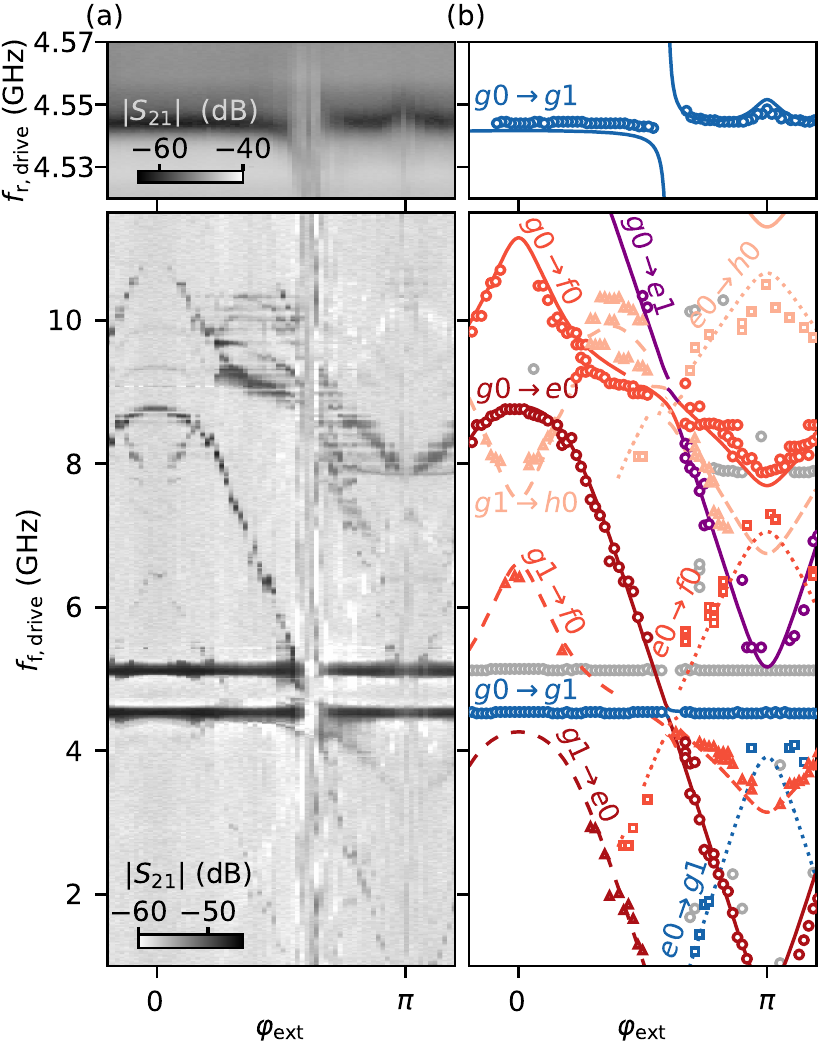}    \caption{{Two-tone spectroscopy of device A, at $B_z=0$.}  (a) Magnitude of the transmitted readout signal as a function of the external flux and $f_{\rm r, drive}$ (top) and $f_{\rm f, drive}$ (bottom), showing the flux modulation of the different transitions. (b) Extrema (maxima and minima) extracted from (a) (markers) and fitted transition frequencies (lines) obtained from the numerical diagonalization of the model Hamiltonian (Methods). Gray markers indicate extrema that are not associated with any fluxonium-resonator transitions. A value of $E_{\rm J}/h = 6.7$ GHz is extracted from the fit.  }
    \label{fig:2}
\end{figure}

We first demonstrate that our device behaves as expected for a fluxonium coupled to a readout resonator. Data from two similar fluxonium devices (device A and device B) will be presented in this article. We first focus on the behavior of device A. We monitored the transmission amplitude, $|S_{21}|$, at frequencies $f_{\rm r, drive}$ around the resonator frequency $f_{g0\rightarrow g1}$, as a function of the external phase $\varphi_{\rm ext} = \frac{2e}{\hbar} \Phi_{\rm ext}$, as shown in the top panel of Fig.~\ref{fig:2}(a). Transitions are labelled as $m_{\rm i}n_{\rm i}\rightarrow m_{\rm e}n_{\rm e}$, where $m_{\rm i}$ ($m_{\rm e}$) and $n_{\rm i}$ ($n_{\rm e}$) are the initial (end) states of the fluxonium and resonator, respectively. The resonator spectrum is periodic in flux and also exhibits gaps in its visibility, which indicate that the resonator is coupled to the fluxonium. The bottom panel in Fig.~\ref{fig:2}(a) shows the flux dependence of  the observed transition frequencies of the fluxonium-resonator system, measured by monitoring the transmission amplitude at $f_{\rm r, drive}=f_{g0\rightarrow g1}$ while the system is driven with a second tone with frequency $f_{\rm f, drive}$, also via the resonator. Though the gradiometric loops that comprise the fluxonium are designed to be symmetric, the placement of the capacitor and gate lines inside them, together with inevitable imprecision when manually depositing the nanowire, lead to a small imbalance between the effective area of the two gradiometer loops. The imbalance allows us to thread a flux using an out-of-plane magnetic field, $B_x$.  Threading a flux quantum through the gradiometer corresponds to $B_x = \SI{550}{\micro T}$, which is much greater than the \SI{15}{\micro T} that would be needed to thread a flux quantum through  one  of  the  two symmetric loops. The gradiometric geometry thus reduces the sensitivity of the fluxonium to magnetic field noise larger than the fluxonium loop by more than an order of magnitude.

To fit the spectroscopy data (markers in Fig.~\ref{fig:2}(b)), we diagonalize the Hamiltonian for the circuit shown in Fig.~\ref{fig:1}(a) \cite{Smith2016}, leaving all circuit parameters free except for $C_{\rm r} =$ \SI{26}{fF}, which we extract from electromagnetic simulations. The parameters obtained from the fit are shown in Tab.~\ref{table:1} and the fitted transition frequencies are denoted with lines in Fig.~\ref{fig:2}(b). Each state is identified by the closest state in energy for the uncoupled system. In addition to transitions originating from the ground state, $g0$, we also observe transitions for which the initial state is $g1$, with one photon in the resonator (dashed lines). This is due to the continuous drive used to monitor $|S_{21}|$ at $f_{g0\rightarrow g1}$, which can populate the resonator. Transitions starting from the first excited fluxonium state, $e0$, around $\varphi_{\rm ext}=\pi$ (dotted lines) are also observed. The transition frequency for $g0 \rightarrow e0$ goes below \SI{1}{GHz} near $\varphi_{\rm ext}=\pi$. The transitions from $e0$ thus occur due to the expected equilibrium thermal occupation of $e0$ for temperatures of around \SI{20}{mK}. We find excellent agreement between the experimental data and the fit, with all fit parameters coming to within $5\%$ of the designed circuit parameters except for \Ej, which we can only coarsely predict. We have etched a \SI{80}{nm}-junction to maximize $E_J$, but the specific values of $E_J$ are determined by the interaction with the electrostatic gate and the microscopic details of the junction.

\begin{table}[h]
\center
\begin{tabular}{ccc}
\hline
\hline
                 & \hspace{0.75cm} Device A \hspace{0.75cm} & Device B  \\
\hline
$E_C/h$ (GHz)    & \hspace{0.5cm}  2.35  \hspace{0.5cm}   &   1.75    \\
$E_L/h$ (GHz)    & \hspace{0.5cm}  0.7  \hspace{0.5cm}   &   1.1    \\
$C_{\rm r}$ (fF) & \hspace{0.5cm}  26  \hspace{0.5cm}   &   26    \\
$L_{\rm r}$ (nH)  & \hspace{0.5cm}   47   \hspace{0.5cm}    &   42    \\
$L_{\rm s}$ (nH)  & \hspace{0.5cm}  8.5     \hspace{0.5cm}    &   4.6     \\
\hline
\hline
\end{tabular}
\caption{Device parameters obtained by fitting spectra in figures \ref{fig:2}, \ref{fig:3} and \ref{fig:4} and by electromagnetic simulations. $h$ is Planck's constant. }
\label{table:1}
\end{table}


\begin{figure*}
\centering
\includegraphics[scale=1.0]{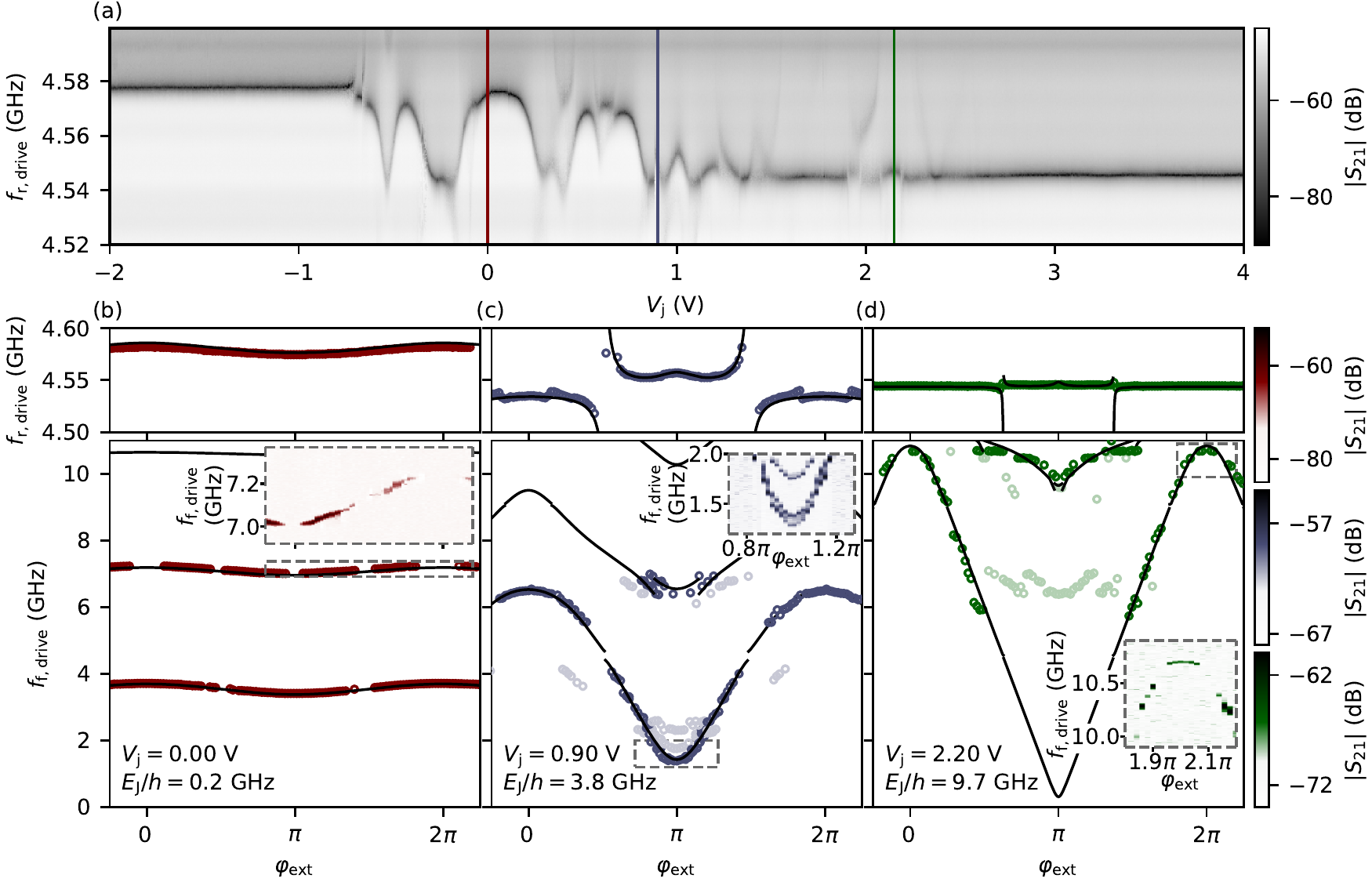}
\caption{{Gate tuning of $E_{\rm J}$ in device A, at $B_z=0$.}  (a) Gate dependence of the resonator's resonant frequency, $f_{g0\rightarrow g1}$, at $\varphi_{\rm ext} = 1.25 \pi$. (b)-(d), Fluxonium spectra at three different gate points, indicated with vertical lines in (a). The markers correspond to the peaks extracted from the measured resonator (top) and two-tone (bottom) transmission data. The fitted transition frequencies  $f_{g0\rightarrow e0}$, $f_{g0\rightarrow f0}$ and $f_{g0\rightarrow h0}$ (black lines) are obtained by fitting the darker markers. The values of $E_{\rm J}/h$ extracted are 0.2, 3.8 and 9.6 GHz respectively. The insets in (b), (c) and (d) show sections of the measured transmission magnitude. In (b) we observe gaps in visibility at zero and half flux in the $g0 \rightarrow f0$ transition. }
\label{fig:3}
\end{figure*}

\section{Electrostatic tuning of fluxonium parameters}
Mesoscopic phenomena often require fine tuning of the charge carrier density in the semiconductor. We here demonstrate that the spectrum of the fluxonium is measurable over a large range of gate voltages and thus does not limit the possible observable phenomena. We first measure $f_{g0\rightarrow g1}$ versus \Vj. As shown in Fig. \ref{fig:3}(a), the resonator frequency is constant at low and high voltage values but has non-monotonic fluctuations in an intermediate range, which is consistent with observations in previous experiments on nanowire junctions \cite{deLange2015, Larsen2015, vanWoerkom2017, Luthi2018}, where these fluctuations were attributed to consecutive openings of different junction channels in the nanowire, whose transparencies oscillate with gate. The behavior of the $f_{g0\rightarrow g1}$ transition provides insight into the \Vj-dependence of the fluxonium \Ej. The value of $f_{g0\rightarrow g1}$ can be seen as the bare resonant frequency of the uncoupled resonator plus a dispersive shift caused by the coupling to the fluxonium. The dispersive shift depends on the frequency of all level transitions of the coupled fluxonium-resonator system and is thus different for different values of \Ej. This change in the dispersive shift leads to the observed changes in the measured resonant frequency of the resonator.

We now investigate directly the fluxonium spectrum. Figures \ref{fig:3}(b-d) present spectra taken at three different \Vj~values (marked by solid lines in Fig.~\ref{fig:3}(a)). Here we use a lower drive power than in Fig.~\ref{fig:2} to reduce broadening of the spectral lines due to the drive power, therefore the main observable transitions start from the ground state, $g0$. For low \Vj, we observe a weakly anharmonic spectrum (Fig.~\ref{fig:3}(b)). For large \Vj~values (Figs.~\ref{fig:3}(c),(d)), however, the flux dependence and the anharmonicity become much stronger. We also note the presence of additional transitions in the spectroscopy data for intermediate and large \Vj~denoted by the lighter markers in Fig.~\ref{fig:3}(c) and (d). The spectra at intermediate \Vj~is taken at a point where the junction is very sensitive to gate voltage; the additional transitions are due to the \Ej~of the junction fluctuating while the spectroscopy is being performed. At higher \Vj, the \Ej~of the junction is stable as a function of gate but additional transitions due to the presence of the resonator drive are also observed. We fit the data of the three spectra using the same parameters as those used in Fig.~\ref{fig:2} while only allowing \Ej~to vary. The fits maintain their accuracy over the whole \Vj~range. We have therefore shown that the behavior of a hybrid fluxonium circuit with a semiconducting weak-link can still be predicted and understood using a simple Hamiltonian with a few degrees of freedom. Moreover, we can conclude that \Ej~is the only circuit parameter affected by large changes in \Vj~and that it generally increases with \Vj. Our results thus show that it is possible to address and observe the state of the system over a large \Ej-range encompassing regimes where its eigenstates are of very different character.

\begin{figure*}
\centering
\includegraphics[scale=1.0]{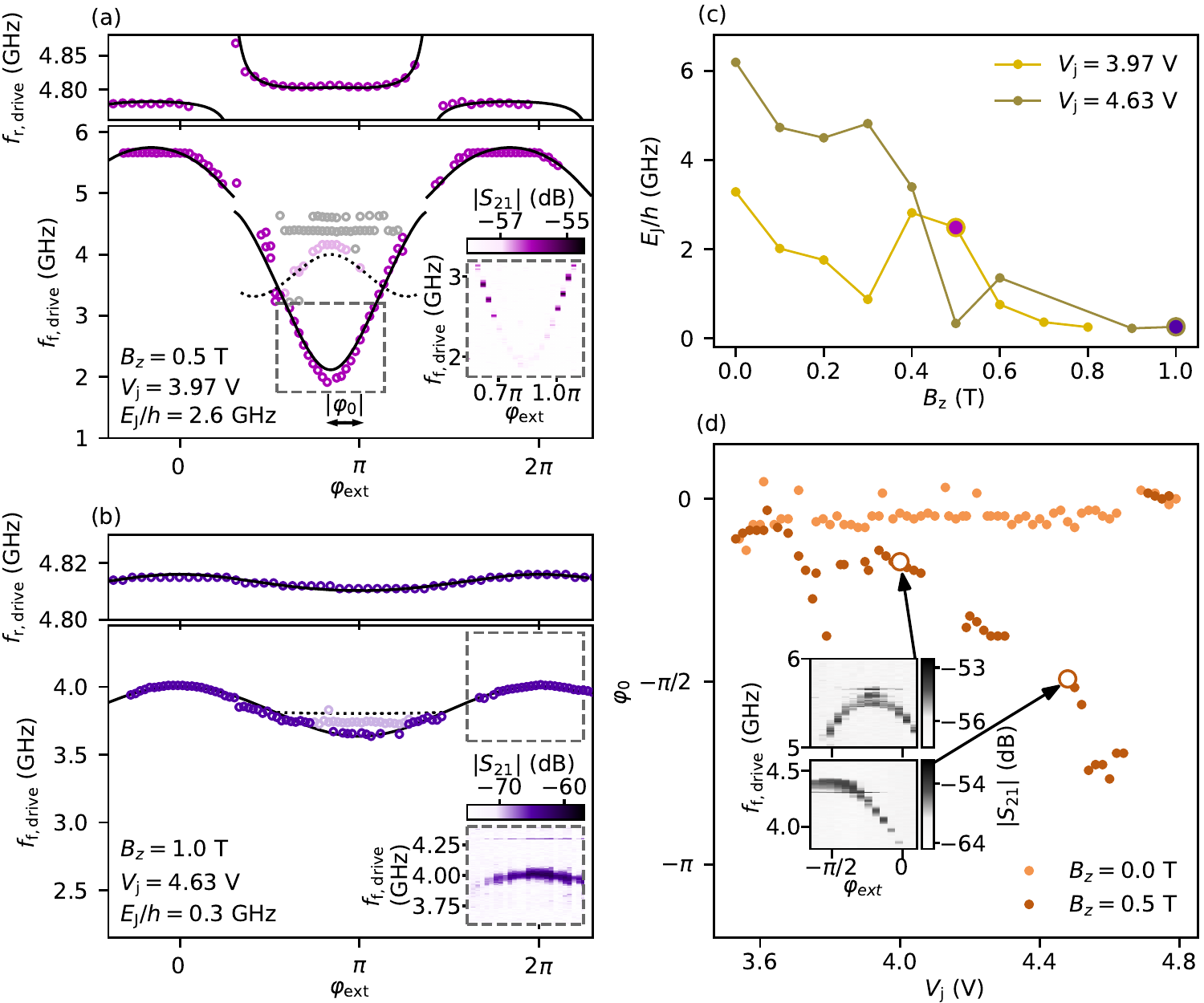}
\caption{{Behavior of device B in magnetic field.} (a), (b),  Fluxonium spectra at two different \Vj~and $B_z$ points. The  \Ej~value extracted from the fit is lower at higher magnetic field. In (a) we observe a $\varphi_0 = -0.16\pi$ phase shift with respect to a reference $\varphi_{\rm ext}$ taken at the same field at $V_{\rm j} = 4.80$ V.  (c) \Ej~versus $B_z$ at two different gate voltage points. \Ej~decreases non-monotonically with field. (d), $\varphi_0$ versus \Vj~at two different magnetic field points. At $B_z=0.0$ T, $\varphi_0$ stays constant for the whole \Vj~range. At $B_z=0.5$ T, however, there is a continuous $\varphi_0$ shift that ranges from 0 to $-\pi$. The two insets show the zero-flux spectroscopy feature shifted from zero at two different gate points. In (c) and (d), the points corresponding to the spectra in (a) and (b) are highlighted with matching colors. }
\label{fig:4}
\end{figure*}

\section{Fluxonium behavior in magnetic field}
Next, we explore the magnetic field compatibility of the nanowire fluxonium. The magnetic field behavior of the device strongly depends on the microscopic details of the nanowire junction. In order to demonstrate the field compatibility of the fluxonium circuit elements, we here show data from device B whose parameters were optimized for magnetic field compatibility. The magnetic field behavior of device A is provided in the Supplementary Material \cite{Supplement}. Spectroscopy measurements at two different \Vj~and $B_z$ points are shown in Figs. \ref{fig:4}(a) and (b). We continue to be able to perform spectroscopy on the fluxonium over the full range in $\varphi_{\rm ext}$ at fields up to \SI{1}{T}. We do note, however, that at higher magnetic fields, the thermal occupation of the excited state of the fluxonium does increase since we observe transitions from this state even when the $g0 \rightarrow e0$ frequency is above \SI{1}{GHz}. Importantly, we can still fit the spectroscopy data accurately in this regime, indicating that the fluxonium-resonator Hamiltonian remains valid at high magnetic fields, with \Ej~being the only parameter largely affected by $B_z$. Fit parameters for device B are shown in Tab. \ref{table:1}.

We finally use the nanowire fluxonium to investigate the behaviour of spin-orbit coupled semiconducting junctions in magnetic field. We perform spectroscopy measurements at gate voltages ranging from 3.5 to \SI{4.8}{V} and fields ranging from 0 to \SI{1}{T}. From the spectroscopy we extract \Ej~as a function of $B_z$ at two different gate points, which is shown in  Fig. \ref{fig:4}(c). A non-monotonic decrease of \Ej~with field is observed at both gate points. We expect an overall decrease in \Ej~versus $B_z$  due to the superconducting gap closing at high magnetic fields. The non-monotonic behaviour of \Ej, however, suggests the presence of interference between different modes in the junction \cite{Zuo2017}. Additional in-field \Ej~data is provided in section IV of the Supplementary Material \cite{Supplement}. An overall decrease of \Ej~with $B_z$ is observed for all investigated gate values, while the non-monotonic dependence is often, but not always, observed.

A shifting of the zero-flux point in the spectroscopy of the fluxonium device at high fields can be used to determine the breaking of multiple symmetries in the semiconducting junction \cite{Yokoyama2014}. This phase shift is known as the $\varphi_0$-Josephson effect, which occurs when chiral and time-reversal symmetries are both broken in the junction. In InSb- and InAs-based junctions, this symmetry-breaking originates from the interplay between the presence of multiple channels in the junction, spin-orbit coupling, and the Zeeman splitting due to the applied magnetic field \cite{Yokoyama2014, Szombati2016, vanWoerkom2017}. We observe such a shift in the zero-flux point of the spectroscopy lines as \Vj~is varied in a $B_z$-field (indicated by $\varphi_0$ in Fig.~\ref{fig:4}(a)). The $\varphi_0$-shift as a function of \Vj~is shown, at $B_z=0$ and at $B_z=\SI{0.5}{T}$, in Fig. \ref{fig:4}(d).The $\varphi_0$ value at $V_{\rm j} = $ \SI{4.80}{V} is taken as the $\varphi_0 = 0$ reference for each $B_z$. The value of $\varphi_0$ is thus a relative value at each $B_z$. At $B_z=0$ the zero-flux point does not change, while it changes continuously with \Vj~when a $B_z$-field is applied. Since the observed phase shift appears as a function of \Vj~at fixed magnetic fields, we can exclude trivial effects such as misalignment of the magnetic field. We note that we observe shifts in $\varphi_0$ approaching $\pi$, which is significantly larger than previously predicted \cite{Yokoyama2014}. Mixing between a large number of spin-split Andreev channels in the junction may lead to the larger observed shifts.

\begin{figure}[h!]
\centering
\includegraphics[scale=1.0]{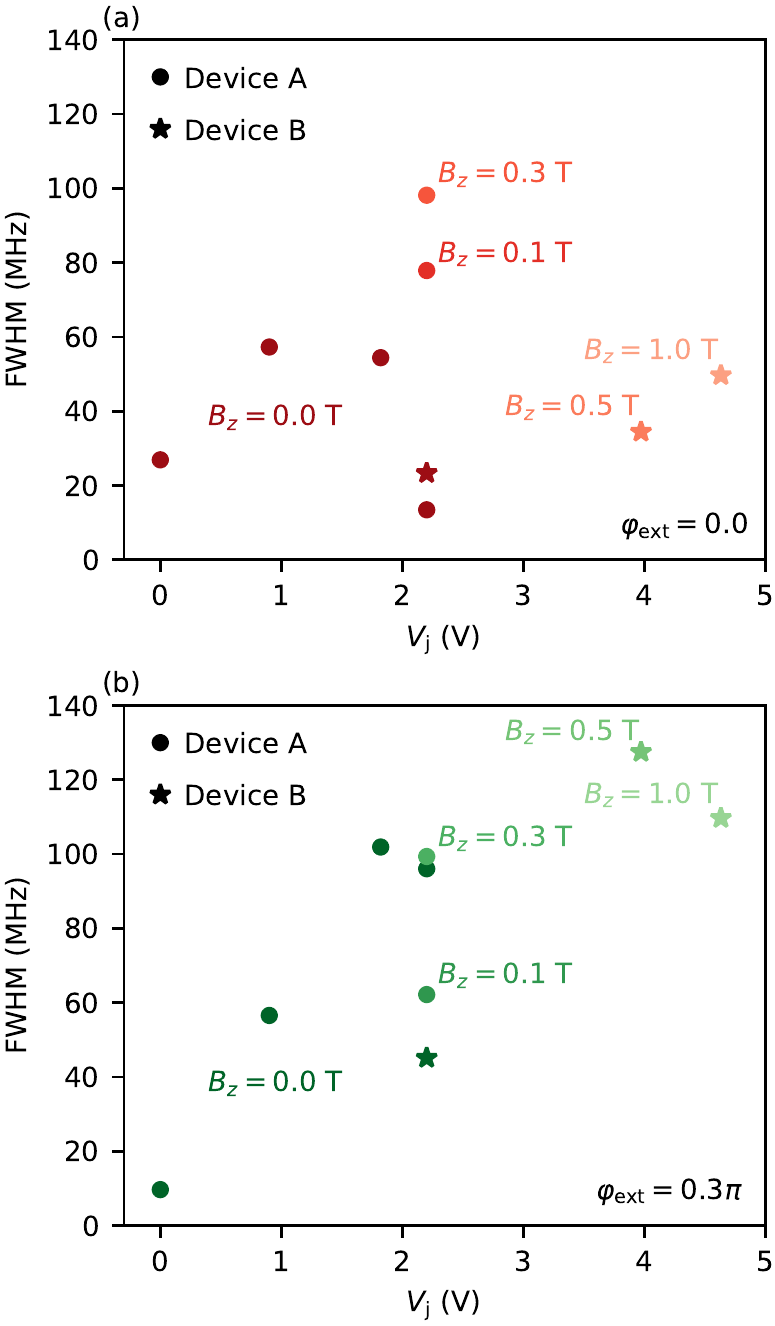}
\caption{{Spectroscopic linewidth.} Extracted linewidth (full width at half maximum (FWHM)) of the $g\rightarrow e$ transition as a function of $V_{\rm j}$ for devices A (circular markers) and B (star markers). Different color strengths denote different magnetic fields. The data for device A corresponds to the spectra shown in Fig.~\ref{fig:3}(b,c,d) and Fig.~3 in the Supplementary Material \cite{Supplement}. The in-field data for device B corresponds to the spectra shown in Fig.~\ref{fig:4}(a,b). The linewidths are extracted in symmetric $0.05\pi$ flux windows around $\varphi_{\rm ext}=0$, which is a sweet spot for the external flux, and $\varphi_{\rm ext}=0.3\pi$ for (a) and (b), respectively.}
\label{fig:5}
\end{figure}

\section{Spectroscopic linewidth}

For the fluxonium's applications as a detector of field-dependent mesoscopic effects, its spectral lines must be measurable under the magnetic field and electrostatic conditions of interest, but also must maintain a narrow linewidth in all regimes. The spectroscopic linewidth bounds the resolution with which we can resolve the splitting of lines due to, for example, quasiparticle poisoning of highly transparent ABSs or MZMs at the junction.

We have performed a spectroscopic linewidth analysis for devices A and B at different points in external flux (Fig.~\ref{fig:5}): near the flux sweet spot at $\varphi_{\rm ext}=0$ and also at $\varphi_{\rm ext}=0.3\pi$. As shown in Fig.~\ref{fig:5}(a), the linewidth of device B stays below 50 MHz up to 1 T at the $\varphi_{\rm ext}=0$. However, the linewidth overall becomes larger, for both devices, at $\varphi_{\rm ext}=0.3\pi$ (Fig.~\ref{fig:5}(b)). This behavior is consistent with residual flux noise broadening the linewidth of the fluxonium. Our linewidth analysis indicates that the energy resolution of the fluxonium would, if measuring near $\varphi_{\rm ext}=0$, be at most 0.4~$\upmu$eV in all regimes, which significantly improves upon the resolution that can currently be reached by transport experiments where the linewidth is limited by thermal broadening (typically $\sim$\SI{10}{\upmu eV}).

We do note that in our experiment, the fluxonium and resonator parameters were chosen for the greatest visibility over the largest range of \Ej~and $\varphi_{\rm ext}$. Improvements upon the design such as weaker coupling to the readout resonator, smaller asymmetry in the gradiometric superinductance, and materials optimizations such as changing the dielectric of the fluxonium capacitor would lead to significantly narrower linewidths.


\section{Conclusion}
In conclusion, we have successfully realized a gate-tunable fluxonium resilient to high magnetic fields. We have combined a gate-controlled junction with magnetic field-compatible materials and a novel gradiometric design to build the hybrid fluxonium. We are able to perform spectroscopy over a large range of gate voltages and in-plane magnetic fields. We have used the fluxonium to investigate the behavior of an InAs Josephson junction in a magnetic field and observed a non-monotonic decrease of the \Ej~of the junction as well as the $\varphi_0$-Josephson effect. The observed $\varphi_0$-shift is gate-tunable up to $\varphi_0=\pi$. One could therefore use this effect to build extremely small, flux-noise insensitive superconducting circuits where the phase difference of the junction could be tuned using a gate-voltage.

The magnetic-field compatible hybrid fluxonium is also now ready to detect the $4\pi$-periodic Josephson effect \cite{Pekker2013} and measure Majorana parity dynamics. We remark that while our experiment was performed with similar materials and in similar magnetic fields to previous experiments where transport signatures pointed to the presence of MZMs, we did not observe any signatures of MZMs. Recent theoretical work has suggested that previously observed signatures of MZMs may be due to trivial Andreev bound states \cite{Huang2018} and has advocated for cleaner materials with stronger spin-orbit coupling proximitized by superconductors with larger energy gaps. Future experiments will thus incorporate InSb nanowires with thinner Al shells as the small junction of the hybrid fluxonium. 

The excellent agreement we have demonstrated between data and theory indicate that one can engineer hybrid circuits such that their behavior can continue to be understood even when the circuit is complicated by the interplay of superconductivity, spin-orbit coupling, and magnetism. This opens avenues for using the hybrid fluxonium to explore superconductivity in the presence of a magnetic field as well as reading out and controlling materials platforms that require the application of a large magnetic field. Since it is straightforward to incorporate different materials into the hybrid fluxonium, proposals to probe superconductor-proximitized edge states in a quantum spin Hall insulator \cite{Dolcini2015, Wu2018} or field-dependent spin-polarized correlated insulating phases \cite{Cao2019, Liu2019} are now possible using our hybrid circuit.

\section*{Acknowledgements}
We thank W. Uilhoorn for fabrication advice. We also thank B. van Heck and A. Antipov for helpful discussions. This research was co-funded by  the allowance for Top consortia for Knowledge and Innovation (TKI’s) from the Dutch Ministry of Economic Affairs and the Microsoft Quantum initiative.

\bibliography{ms}

\end{document}


\title{Supplementary information for ``A gate-tunable, field-compatible fluxonium"}

\author{Marta Pita-Vidal}
\affiliation{QuTech, Delft University of Technology, 2628 CJ Delft, The Netherlands.}
\affiliation{Kavli Institute for Nanoscience, Delft University of Technology, 2628 CJ Delft, The Netherlands.}

\author{Arno Bargerbos}
\affiliation{QuTech, Delft University of Technology, 2628 CJ Delft, The Netherlands.}
\affiliation{Kavli Institute for Nanoscience, Delft University of Technology, 2628 CJ Delft, The Netherlands.}

\author{Chung-Kai Yang}
\affiliation{Microsoft Quantum Lab Delft, 2628 CJ Delft, The Netherlands.}

\author{David J. van Woerkom}
\affiliation{Microsoft Quantum Lab Delft, 2628 CJ Delft, The Netherlands.}

\author{Wolfgang Pfaff}
\affiliation{Microsoft Quantum Lab Delft, 2628 CJ Delft, The Netherlands.}

\author{Nadia Haider}
\affiliation{QuTech, Delft University of Technology, 2628 CJ Delft, The Netherlands.}
\affiliation{Netherlands Organisation for Applied Scientific Research (TNO), 2628 CK Delft, The Netherlands}

\author{Peter Krogstrup}
\affiliation{Center for Quantum Devices, Niels Bohr Institute, University of Copenhagen, 2100 Copenhagen, Denmark and  Microsoft Quantum Materials Lab Copenhagen, 2800 Lyngby, Denmark.}

\author{Leo P. Kouwenhoven}
\affiliation{QuTech, Delft University of Technology, 2628 CJ Delft, The Netherlands.}
\affiliation{Microsoft Quantum Lab Delft, 2628 CJ Delft, The Netherlands.}

\author{Gijs de Lange}
\affiliation{Microsoft Quantum Lab Delft, 2628 CJ Delft, The Netherlands.}

\author{Angela Kou}
\affiliation{Microsoft Quantum Lab Delft, 2628 CJ Delft, The Netherlands.}


\date{\today}

\maketitle
\section{Fitting procedure}

To find the relative extrema we apply a peak finding algorithm to the raw data. This algorithm first smooths the data in the frequency axis to avoid errors in peak finding due to noise. A minimum peak height is specified. 

We fit the extracted data with the Hamiltonian corresponding to the circuit model in Fig.~1(a) in the main text. The fluxonium Hamiltonian, $\hat{H}_{\rm f}$, and the total Hamiltonian of the coupled readout-fluxonium system, $\hat{H}$, can be written in terms of two degrees of freedom, $\hat{\varphi}_{\rm f}$ and $\hat{\varphi}_{\rm r}$ (the phase drops across the fluxonium junction and across $C_{\rm r}$, respectively), and their conjugated charges $\hat{n}_{\rm f}$ and $\hat{n}_{\rm r}$ \cite{Smith2016}. In the limit $L_{\rm f} \gg L_{\rm s}, L_{\rm r}$ (where $L_{\rm f} = \frac{\Phi_0^2}{4\pi^2E_L}$ and $\Phi_0=h/2e$),
\begin{equation} \label{eq:Hf}
\hat{H}_{\rm f}= 4 E_C\hat{n}_{\rm f}^2 - E_{\rm J}(V_{\rm j}, B){\rm cos}(\hat{\varphi}_{\rm f}) + \frac{1}{2}E_L\big(\hat{\varphi}_{\rm f}-\varphi_{\rm{ext}}\big)^2
\end{equation}
and
\begin{equation} \label{eq:H}
\hat{H} = \frac{2e^2}{C_{\rm r} }\hat{n}_{\rm r}^2 + \frac{1}{2}\frac{(\Phi_0/2\pi)^2}{(L_{\rm r}+L_{\rm s}) }\hat{\varphi}_{\rm r}^2 - \frac{1}{2}\frac{(\Phi_0/2\pi)^2L_{\rm s}}{L_{\rm f} (L_{\rm r} + L_{\rm s})}\hat{\varphi}_{\rm r}\hat{\varphi}_{\rm f}  + \hat{H}_{\rm f}.
\end{equation}
The first two terms of $\hat{H}$ describe the uncoupled resonator, while the third term accounts for the coupling between resonator and fluxonium.

We diagonalize Hamiltonian \ref{eq:H} using the numerical method in \cite{Smith2016}. All the spectra for the same device are fit simultaneously. The free parameters $E_C$, $E_L$, $L_{\rm r}$ and $L_{\rm s}$ are common for all spectra corresponding to the same device. The free parameter $E_{\rm J}$, however, has a different value for each spectrum. 

All markers shown in Figs. 2-4 in the main text are included in the fits. The marker colors are assigned by association to the different transitions. Gray markers indicate extrema that could not be assigned to any transition included in the fit.

\section{Theoretical model for the uncoupled fluxonium}

\begin{figure}[hp!]
    \centering
    \includegraphics[scale=1.0]{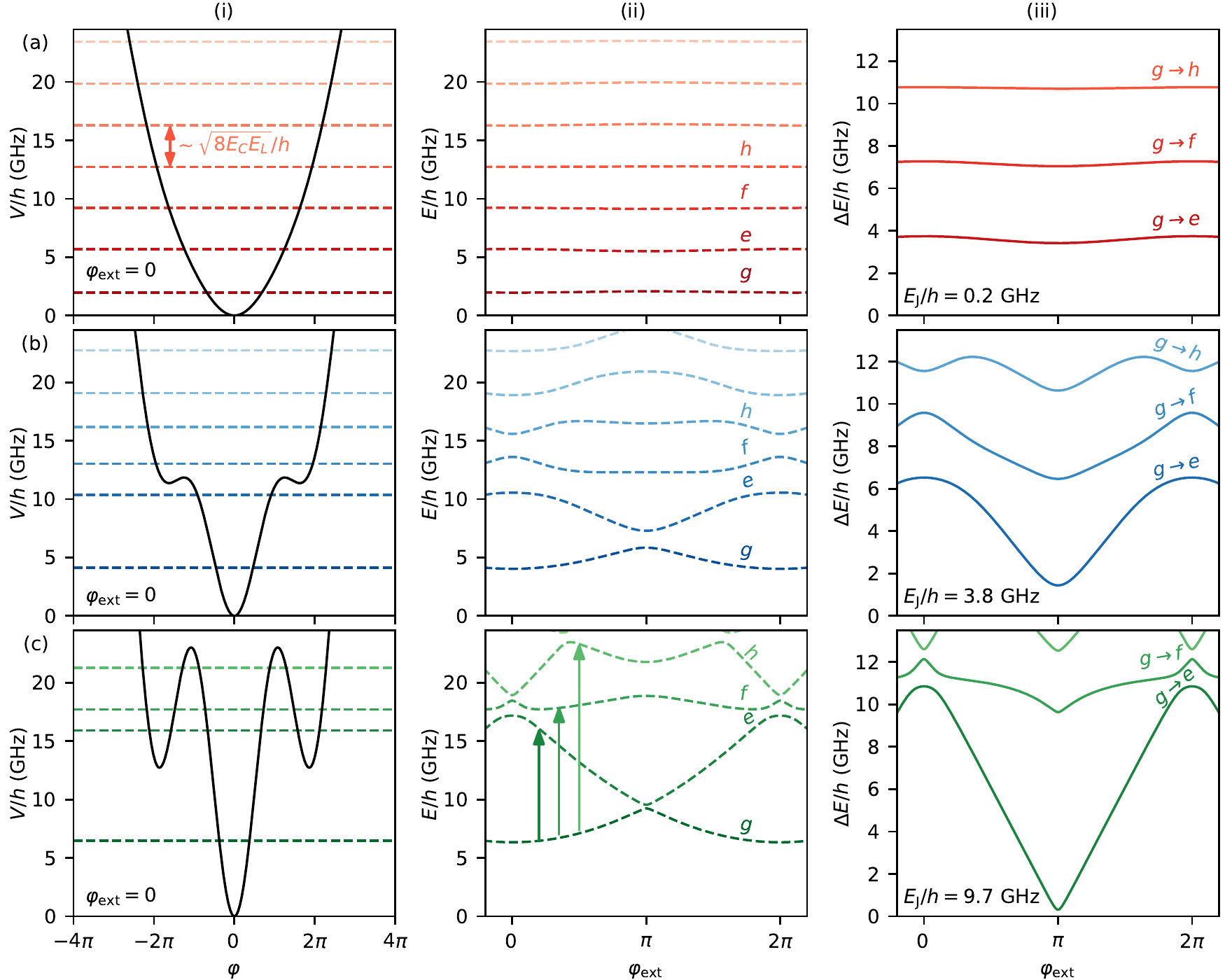}
    \caption{{Potential, energy spectrum and transition frequencies of the uncoupled fluxonium.} Rows (a), (b) and (c) correspond to  $E_{\rm J}/h = $ 0.2, 3.8 and 9.6 GHz, respectively, for the model parameters of device A. Column (i) shows the fluxonium potential in equation \ref{eq:V}, at $\varphi_{\rm ext}=0$, in black lines. The lowest eigenenergies of Hamiltonian \ref{eq:Hf}  at $\varphi_{\rm ext}=0$ are superimposed as horizontal dashed lines. Column (ii) shows how these energies disperse as $\varphi_{\rm ext}$ is varied. The solid arrows indicate transitions starting from the ground state, $g$. The energies of these transitions are plotted in column (iii) as a function of $\varphi_{\rm ext}$.}
    \label{fig:s1}
\end{figure}

The Hamiltonian for the uncoupled fluxonium, $\hat{H}_{\rm f}$, can be written in terms of the phase drop across the junction, $\hat{\varphi}_{\rm f}$, and its conjugate charge $\hat{n}_{\rm f}$ \cite{Manucharyan2009, Smith2016}
\begin{equation} \label{eq:Hf}
\hat{H}_{\rm f}= 4 E_C\hat{n}_{\rm f}^2 - E_{\rm J}(V_{\rm j}, B){\rm cos}(\hat{\varphi}_{\rm f}) + \frac{1}{2}E_L\big(\hat{\varphi}_{\rm f}-\varphi_{\rm{ext}}\big)^2.
\end{equation}
Each of the terms in $\hat{H}_{\rm f}$ results from one of the three characteristic energies of fluxonium: $E_C$, $E_L$ and $E_{\rm J}$. The two conjugate variables in this Hamiltonian, $\hat{\varphi}_{\rm f}$  and $\hat{n}_{\rm f}$, are analogous to position and momentum, respectively. With this interpretation, the two terms involving $\hat{\varphi}_{\rm f}$  constitute a phase-dependent potential 
\begin{equation} \label{eq:V}
V(\varphi)=  - E_{\rm J}{\rm cos}(\varphi) + \frac{1}{2}E_L\big(\varphi-\varphi_{\rm{ext}}\big)^2.
\end{equation}
Fig. \ref{fig:s1}(i) shows the potential at $\varphi_{\rm{ext}}=0$ for the three $E_{\rm J}$ values in Fig. 3 in the main text. The $E_L$ term results in a  parabolic background, common for the three cases. The $E_{\rm J}$ adds a periodic modulation on top of it, which becomes more noticeable as $E_{\rm J}$ increases. $E_C$ can be seen as a mass term and, together with the potential $V$, determines the eigenstates of fluxonium. The lowest energy eigenstates are labelled  $g$, $e$, $f$ and $h$. Their energies at $\varphi_{\rm{ext}}=0$ are shown as colored lines in Fig. \ref{fig:s1}(i).  Fig. \ref{fig:s1}(ii) shows the energy dispersion of the fluxonium with respect to $\varphi_{\rm{ext}}$. For small $E_{\rm J}$ values, the variation of $\varphi_{\rm{ext}}$ results in weak oscillations of the eigenenergies, while for large $E_{\rm J}$'s the $\varphi_{\rm{ext}}$-dependence is much stronger. In the limit of small $E_{\rm J}$ the eigenstates are vibrational modes of the  harmonic LC oscillator determined by $E_L$ and $E_C$. Therefore, their energies are evenly spaced with a separation determined by the plasma frequency $\sqrt{8E_CE_L}/h$. In the limit of large $E_{\rm J}$ the eigenstates become superpositions of persistent currents localized in phase.

Fig. \ref{fig:s1}iii shows the transition frequencies between different pairs of eigenstates, which are the quantities that can be addressed experimentally. For simplicity, only the transitions starting from the ground state are shown here. The two-tone spectroscopy data in Figures 2, 3 and 4 in the main text shows the measured transition frequencies for the coupled fluxonium-resonator system.  

\clearpage
\section{Modeling the junction accounting for highly transparent transmission channels}

In the main text the spectroscopy data is fit assuming a sinusoidal current-phase relation at the junction, which results in the term $-E_{\rm J} {\rm cos}(\varphi)$ in the Hamiltonian potential (equation \ref{eq:V}), which is characteristic of superconductor-insulator-superconductor (SIS) junctions with low transparency channels. A more accurate model for the potential of a semiconducting Josephson junction with $N$ channels with transparencies $T_i$ is \cite{Beenakker1991}
\begin{equation} \label{eq:Tpotential}
V_{\rm sJJ}(\varphi)=\Delta \sum_{i=1}^N \sqrt{1-T_i{\rm sin}^2(\varphi/2)},
\end{equation}
where $\Delta$ denotes the induced gap on the proximitized sections of the nanowire. When the transparencies of the different channels are low, this dependence converges to a sinusoidal relation with $E_{\rm J}=\Delta \sum_{i=1}^N T_i/4$.\\

\begin{figure}[hp!]
    \centering
    \includegraphics[scale=1.0]{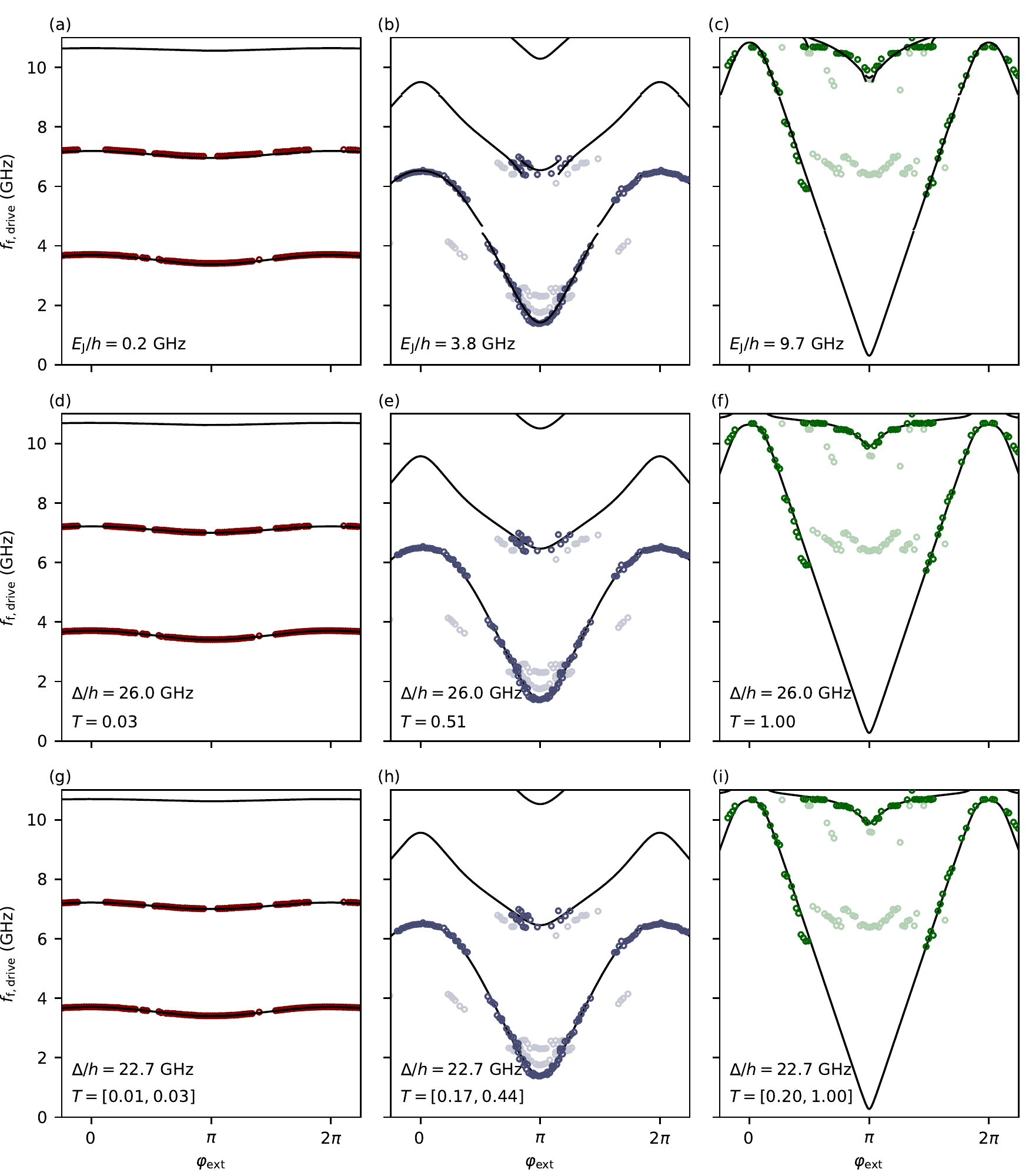}
    \caption{{Fits using the Andreev potential in the Hamiltonian.} Spectroscopy data fit with the uncoupled fluxonium Hamiltonian (equation \ref{eq:Hf}) with a sinusoidal potential (first row), with a one-channel potential (second row) and with a two-channel potential (third row). Only the dark markers, corresponding to transitions starting from the ground state, are included in the fit. $E_C/h=2.35$~GHz and $E_L/h=0.7$~GHz are obtained for the three different fits. The rest of the fit parameters are indicated in the different panels.}
    \label{fig:s5-nores}
\end{figure}

Here we show fits to the spectroscopy data from Fig.~3 in the main text using the semiconducting junction potential $V_{\rm sJJ}$ with different number of channels. Fig.~\ref{fig:s5-nores} shows fits to the transitions starting from the ground state with the Hamiltonian for the uncoupled fluxonium (equation \ref{eq:Hf}) using a sinusoidal potential, a one-channel potential and a two-channel potential. For low and intermediate $E_{\rm J}$ values the accuracy of the different models is very similar. For high $E_{\rm J}$, however, the fits to the high frequency transitions are more accurate when highly transparent channels are included (Fig.~\ref{fig:s5-nores} third column). This points toward the presence of at least one highly transparent channel in the junction. The fit accuracy is however similar when one, two or more (not shown) channels are considered. We thus can not extract a measure of the number of junction channels by performing fits of the spectroscopic data. \\

\begin{figure}[hp!]
    \centering
    \includegraphics[scale=1.0]{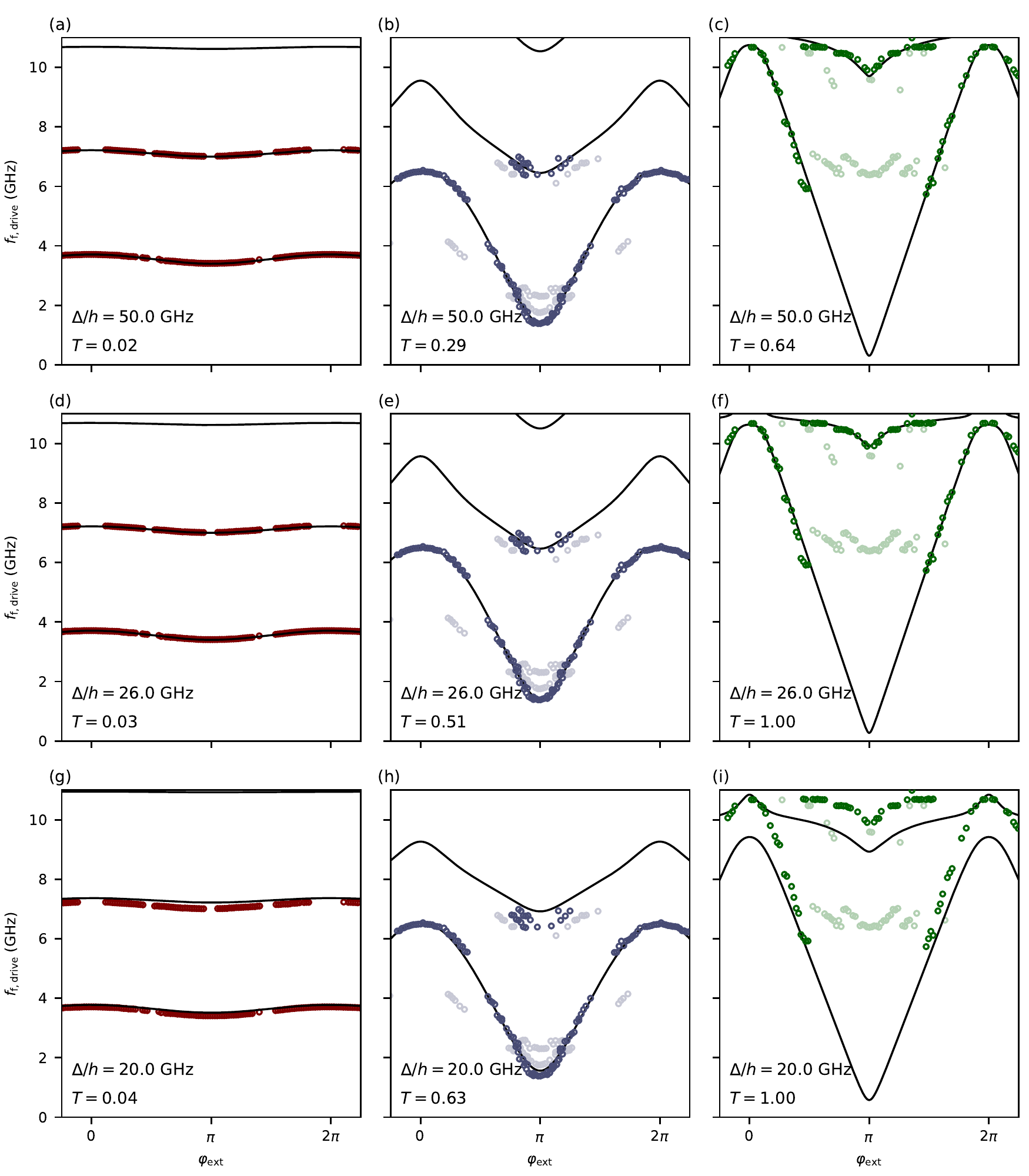}
    \caption{{$\Delta$-dependence of fits using the Andreev potential in the Hamiltonian} Spectroscopy data  fitted with the uncoupled fluxonium Hamiltonian (equation \ref{eq:Hf}) with a one-channel potential. In the first and third rows the value of $\Delta$ is fixed to $\Delta/h=50.0$~GHz and $\Delta/h=20.0$~GHz, respectively. In the second row $\Delta$ is left free and $\Delta/h=26.0$~GHz is obtained as the optimum value. Only the dark markers, corresponding to transitions starting from the ground state, are included in the fit. }
    \label{fig:s5-delta}
\end{figure}

Fig.~\ref{fig:s5-delta} shows the $\Delta$-sensitivity of the one-channel spectrum, showing the fit accuracy for $\Delta$-values lower and higher than the optimum value. We note that the best fit assuming a single channel occurs at $\Delta=$~\SI{26}{GHz}. This is smaller than the gap ($\Delta=$~\SI{53}{GHz}) typically measured in transport experiments. This discrepancy may be due to the necessity of incorporating excited Andreev states into the fluxonium model or interactions due to an accidental quantum dot in the junction \cite{Hazard2019, Bargerbos2019, Kringhoj2019}.\\

\clearpage
\section{Resonator-fluxonium inductive coupling for device B}

\begin{figure}[hp!]
    \centering
    \includegraphics[scale=1.0]{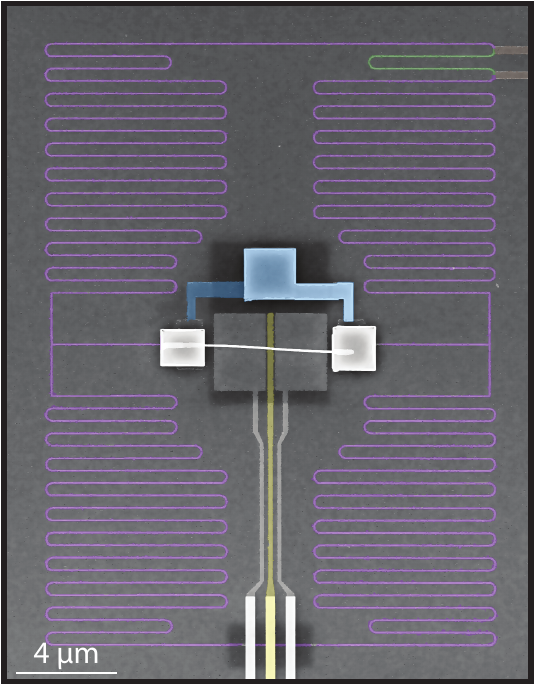}
    \caption{False-colored SEM image showing the resonator-fluxonium inductive coupling for a device with design identical to the one of device B. Parts implementing the different circuit elements are shaded with the same colors as in Fig.~1 in the main text.}
    \label{fig:s12}
\end{figure}

The SEM image shown in Fig.~1(d) in the main text corresponds to a device with design identical to the one of device A, for which the shared inductance is situated on the top part of the fluxonium loop. For device B, however, the shared inductance is placed on a side of the superinductive loop, as shown in Fig.~\ref{fig:s12}. We do not expect this difference in design to have had a measurable impact on the measurements of Device A and B.\\

\clearpage
\section{Field data for device A}

\begin{figure}[hp!]
    \centering
    \includegraphics[scale=1.0]{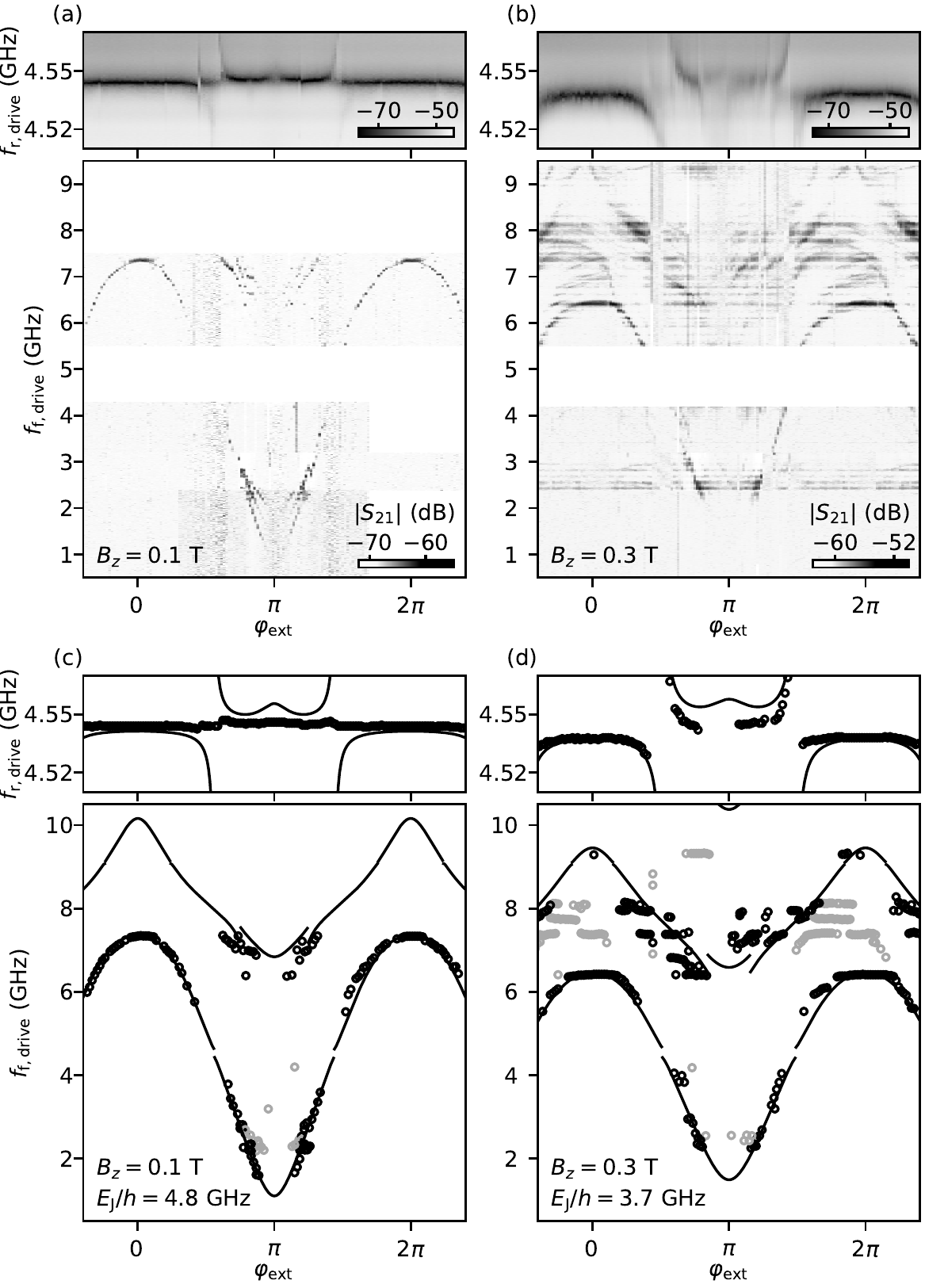}
    \caption{{Spectroscopy data for device A under in-plane magnetic field.} Spectra in (a) and (b) are taken at $B_z = $ \SI{0.1}{T} and \SI{0.3}{T}, respectively, at the same gate voltage, $V_{\rm j}=$ \SI{2.20}{V}, as the spectrum in Fig. 3c in the main text. (c)  and  (d) show the extracted peaks (markers) and the fitted transition frequencies (solid lines). The $E_{\rm J}/h$ values obtained from the fit are 4.8 and 3.7 GHz, respectively.}
    \label{fig:s3}
\end{figure}

Fig. \ref{fig:s3} shows the spectroscopy data for device A under in-plane magnetic field, up to \SI{0.3}{T}. We note the presence of phase-independent lines crossing the fluxonium spectrum at frequencies above \SI{6}{GHz} at \SI{0.3}{T}. Beyond \SI{0.3}{T}, the transition frequencies of the fluxonium came within \SI{1}{GHz} of the resonator frequency, which resulted in the spectroscopy of the fluxonium becoming unmeasureable.

\clearpage
\section{Additional field and gate data for device B}

\begin{figure}[hp!]
    \centering
    \includegraphics[scale=1.0]{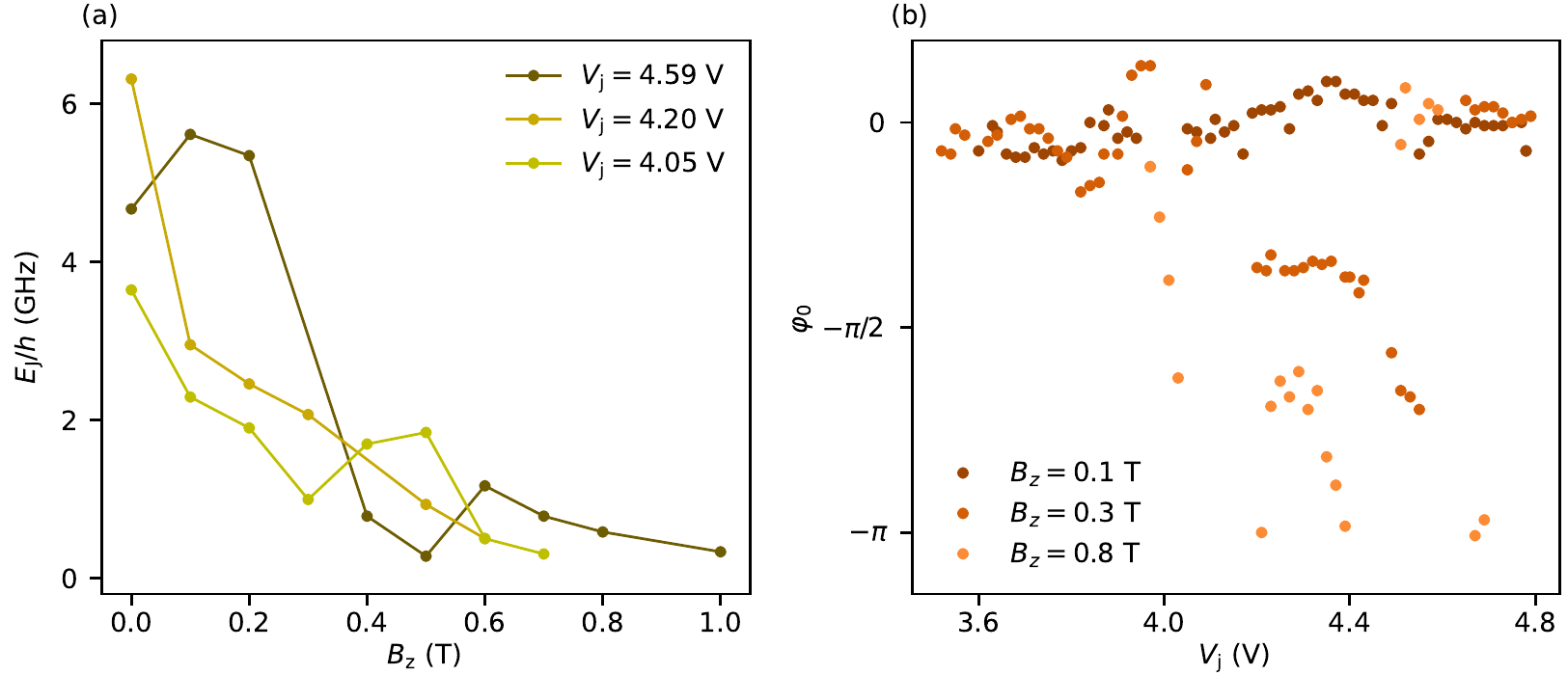}
    \caption{{Extra data for device B in magnetic field.}  (a), $E_{\rm J}$ versus $B_z$ at three different $V_{\rm j}$ points that complement the data shown in Fig.~4(c) in the main text.  (b), $\varphi_0$ versus $V_{\rm j}$ at three different $B_z$ points that complement the data shown in Fig.~4(d) in the main text. $V_{\rm j}=4.8$~V is taken as the  $\varphi_0=0$ reference for $B_z=0.1$~T and $B_z=0.3$~T and $V_{\rm j}=4.6$~V is taken as the $\varphi_0=0$ reference for $B_z=0.8$~T.}
    \label{fig:s4}
\end{figure}

Fig.~\ref{fig:s4} shows $E_{\rm J}$ and $\varphi_0$ extracted from spectroscopy data at multiple field and gate values, complementing the data shown in Fig.~4(c,d) in the main text. The Josephson energy decreases with $B_z$ for all $V_{\rm j}$ values (Fig.~\ref{fig:s4}(a)) and does so in a non-monotonic way for multiple gate points. We observe the anomalous Josephson effect starting from \SI{0.1}{T} and becoming stronger for larger values of $B_z$ (Fig.~\ref{fig:s4}(b)). Above \SI{0.8}{T}, $E_{\rm J}$ is very low for multiple gate voltages, which makes it difficult to observe the $\varphi_0$-shift at those points.

\bibliography{supplement}